\documentstyle [12pt] {article}
\begin {document}

\begin {center}
{ \large \bf GRAVITATIONAL FIELDS AND
             DARK MATTER}
\vspace {2mm}

             {\large  Valery Koryukin} \\
\vspace {2mm}

{ \it Mari State Technical University, 
 sq. Lenin 3, Yoshkar--Ola, 424024, Russia} \\
 e-mail: koryukin@marstu.mari.ru

\end {center}

\vspace {10mm}
 In this paper a hypothesis is considered, in which neutrinos
 and other weakly interacting particles play a fundamental role
 in Universe. In addition the Newton gravitational constant~$G_N$
 and the Hubble constant~$H$ are interpreted as parameters,
 characterizing the neutrinos background of Universe.

\vspace{10mm}

 In 1925 Shirokov~\cite {shi} has shown, that the solution of
 the Laplace equation dependent only on spacing interval~$r$
 between points and possible only, when the space curvature is
 a constant one, has the form: $V(r) = (c_1/R) \cot(r/R) + c_2$,
 if the curvature is the positive one ($R$ is a radius of a curvature;
 $c_1, c_2$ are the arbitrary constants) or~\cite {che}
\begin{equation}
\label{1}
 V(r) = (c_1/L) \coth (r/L) + c_2 ,
\end{equation}
 when the curvature of space is negative ($L$ is the Lobachevskij
 constant). For obtaining a potential of Newton (or Coulomb)
 It is necessary to put $c_2 =-c_1/L $, $L\to\infty $
 ($c_2 =-c_1/R $, $R\to\infty $). Let's consider hereinafter,
 that at availability of the space curvature to be obliged non-zero,
 owing to what the Newton potential should be exchanged by a
 potential (1), in which $c_2 =-c_1/L$ And to which one actually
 has come Lobachevskij~\cite{che}, studying properties of the spaces
 with negative curvature. For such potential misses the Seeliger
 paradox~\cite{see} (for the matter uniformly distributed in the
 space and with
 the Newton law a full potential is divergenting). For the Lobachevskij
 potential, as it is uneasy to note, the divergence misses
 (the relativistic generalization of the Lobachevskij theory on the
 basis the Einstein theory was proposed by Chernikov~\cite{che}).

 Allowing quantum nature of exchange of quasi-particles, we shall record
 the Lobachevskij gravity potential in the form:
\begin{equation}
\label{2}
 V(r)=\frac{A}{L} (1- \coth{\frac{r}{L}})
  =-\frac{2A}{L}~\frac{e^{-2r/L}}{1-e^{-2r/L}} 
 =-\frac{2A}{L}~\sum_{n=1}^{\infty} e^{-2rn/L},
\end{equation}
 where~$A = G_N m_1 m_2 $
 ($G_N\approx 6.7\cdot 10^{-39} GeV^{-2} $ is gravitational the
 Newton constant; the system of units will hereinafter be used
 $h/(2\pi) =c=1$, where $h$ is the Planck constant, and $c$ is
 the light speed; $m_1, m_2$ are the masses of interacting bodies).
 Let's remark, that the asymptotical behavior of the Lobachevskij
 potential will be reduced to the behavior of the potential
\begin{equation}
\label{3}
 V(r) = - (A/r) e^{-Br},
\end {equation}
 which, on our opinion, for the first time was offered by
 Neumann~\cite {neu} and which is more known as the Yukawa
 potential, introduced for the description of short-range
 nuclear forces (the similar potential will be used and for
 the description of short-range electromagnetic fields in a
 plasma). As a result constant $B=1/L$ in gravity potential
 (3), as well as $L$ in a potential (2), should be determined
 by properties of a medium (the dark matter), in addition according
 to our reckoning the fundamental role
 should play the background neutrinos of the Universe.
 It is possible to assume, that the constant $B=1/L $ will be
 to coincide value of the Hubble constant
 $H\approx 1.134\cdot 10^{-42}GeV$ (or even
 $B\propto H $), which one in
 this case will serve one of the characteristics of
 the dark matter (or, it is more concrete, of the neutrinos medium).

 We shall section a matter of the Universe
 on the rapid subsystem and the slow one, considering, that all
 known particles (it is possible, excluding only neutrino) belong
 to the rapid subsystem also are described by standard fields of
 the quantum field theory. Considering fundamental particles as
 coherent frames in open systems, characterizing by a quasi-group
 structure, we shall use inhomogeneous (quasi-homogeneous)
 space-time manifold allotted by the geometrical structure
 of the Riemannian space.
 In same time for the description of the slow subsystem
 (weakly interacting particles) we shall apply
 the condensed description through
 mixtures of gauge fields
 having non-zero vacuum averages~\cite {kor}
 (in particular it is convenient to use
 fields $\Phi_i^{(k)} (x), \Phi^j_{(l)} (x)$~\cite{tre};
 indexes $i, j, k, l,... $ and $i, j, k, l,... $
 receive values $ 1,2,3,4 $; a point $x\in M_4 $, where $M_4 $ is
 the space-time manifold; $\Phi_i^{(k)} \Phi^j_{(k)} =\delta_i^j$,
 $\Phi_i^{(k)} \Phi_j^{(l)} \eta_{(k)(l)} =g_{ij}$,
 $\delta_i^j$ are the Kronecker delta symbols,
 $\eta_{(k)(l)}$ are the covariant components of the metric tensor
 of the Minkowski space,
 $g_{ij}$ are the covariant components of the metric tensor of the
 Riemannian space-time), using them as gravity potentials.
 For obtaining the Einstein gravitational equations
 a full Lagrangian ${\cal L}_t$ let's write to a view
$$
 {\cal L}_t = {\cal L} (\Psi, D\Psi) + 
 \eta^{(j)(m)}~[\kappa_o~F_{(i)(j)}^{\underline{a}}~
 F_{(k)(m)}^{\underline{b}}~\eta^{(i)(k)}
 ~\eta_{\underline{a}\underline{b}} +
$$
\begin{equation}
\label{4}
 \kappa_1~(F_{(i)(j)}^{(k)}
 F_{(l)(m)}^{(n)}~\eta^{(i)(l)}~\eta_{(k)(n)} +
 2~F_{(i)(j)}^{(k)}~F_{(k)(m)}^{(i)} -
 4~F_{(i)(j)}^{(i)}~F_{(k)(m)}^{(k)})]/4
\end {equation}
 ($\underline{a}, \underline{b}, \underline{c}, \underline{d},
 \underline{e} = 5, 6, ..., 4+\underline{r}$),
 where $\eta_{\underline{a}\underline{b}}$
 are the covariant components of the metric tensor
 of the flat space,
$\kappa_o = 1/(4\pi)$ and $\kappa_1 = 1/(4\pi G_N)$
 ($G_N\approx 6.7\cdot 10^{-39} GeV^{-2} $ is gravitational the
 Newton constant; the system of units will hereinafter be used
 $h/(2\pi) =c=1$, where $h$ is the Planck constant, and $c$ is
 the light speed),
\begin{equation}
\label{5}
 F_{(i)(j)}^{(k)}
 = (\Phi_{(i)} ^m\nabla_m\Phi_{(j)}^l-
  \Phi_{(j)}^m\nabla_m\Phi_{(i)}^l)~\Phi_l^{(k)}+ 
 A_{(i)}^{\underline{a}}~T_{\underline{a}}{}_{(j)}^{(k)} -
 A_{(j)}^{\underline{a}}~T_{\underline{a}}{}_{(i)}^{(k)}
\end {equation}
 are the components of the intensities
 of the gravitational fields~$\Phi_{(i)}^k (x)$,
\begin{equation}
\label{6}
 E_{(i)(j)}^{\underline{a}} =
 \Phi_{(i)}^k~\Phi_{(j)}^l~(\nabla_k A_l^{\underline{a}}
 -\nabla_l A_k^{\underline{a}}+
 A_k^{\underline{b}}~A_l^{\underline{c}}
 ~C_{\underline{b}\underline{c}}^{\underline{a}}
 +C_{k\underline{b}}^{\underline{a}}~A_l^{\underline{b}}-
 C_{l\underline{b}}^{\underline{a}}~A_k^{\underline{b}}
 +C_{kl}^{\underline{a}})
\end {equation}
 are the components of the intensities
 of fields~$A_i^{\underline{a}}(x)$,
 $\Psi (x)$ are the fields describing fermions ($\nabla_k$
 are covariant derivatives).
 It allows to connect the constant
 $\kappa_1\propto 1/G_N $ with the density of particles
 (quasiparticles) of the slow subsystem of the Universe and to
 consider a small magnitude of the gravitational constant~$G_N$
 be a consequent of a large density of particles (quasi-particles) of
 the slow subsystem described by fields~$\Phi_i^{(k)} (x)$,
 $\Phi^j_{(l)} (x)$, and to rewrite down the Lagrangian (4)
 in the form
$$
 {\cal L}_t = {\cal L}(\Psi, D\Psi) +
 \kappa~\{F_{\alpha \beta}^{\gamma}~F_{\delta \kappa}^{\epsilon}~
 [\eta^{\alpha \delta}~(\delta_{\gamma}^{\kappa}
 ~\delta_{\epsilon}^{\beta} -
 2~\delta_{\gamma}^{\beta}~\delta_{\epsilon}^{\kappa}) +
 \eta^{\beta \kappa}~(\delta^{\alpha}_{\epsilon}~
 \delta_{\gamma}^{\delta}
 - 2~\delta^{\alpha}_{\gamma}~\delta^{\delta}_{\epsilon}) +
$$
\begin{equation}
\label{7}
 \eta_{\gamma \epsilon}~(\eta^{\alpha \delta}~\eta^{\beta \kappa}
 - 2~\eta^{\alpha \beta}~\eta^{\delta \kappa})]
 + F_{\alpha \beta}^{\underline{a}}~
 F_{\delta \kappa}^{\underline{b}}~
 \eta_{\underline{a}\underline{b}}~
 (\eta^{\alpha \delta}~\eta^{\beta \kappa}
 - 2~\eta^{\alpha \beta}~\eta^{\delta \kappa})\}/4
\end{equation}
 (the cardinality of the values set of the Greek indexes
 is equal to ${\cal N}$),
 where $\eta_{\alpha\beta}$ are the covariant components of the metric
 tensor of the flat space,
 $\eta_{\alpha\beta}~\eta^{\alpha\gamma}=\delta_{\alpha}^{\gamma}$,
 $\kappa$ is constant, and the generalization of the
 formulae (5), (6) is obvious.

 Let's consider, that violation of a symmetry in the weak interactions
 is induced by a high density of right-handed polarized
 neutrinos of different flavors and,
 accordingly, left-handed polarized antineutrinos, which at
 low energies do not participate in reactions owing to the large
 pressure in matching degenerate Fermi --- gases.
 In addition
 it is necessary to recall the Dirac hypothesis of 1930, in which
 the dilemma (put by existence of the solutions, offered him
 equations) is resolved by filling by electrons all
 states with negative energies pursuant to a principle of
 the Pauli prohibition.
 In outcome a state of vacuum is identifiable as state, in which
 all levels with negative energies are filled by particles
 (weakly interacting particles), and all levels with positive
 energies are free, that corresponds to the completely
 degenerate Fermi --- gas at the zero temperature.
 The slightest increase of a temperature, which can be and
 by consequent of a fluctuation (here again it is necessary
 to recollect a Boltzman hypothesis,
 asserting, that observed by us
 the area of the Universe is outcome of a huge fluctuation) will cause
 the appearance of excited states --- known elementary particles
 with positive energy having colour and
 (or only) electrical charges.
 If in addition the space-division of weakly interacting particles
 is descending, we shall receive the charge
 asymmetrical Universe with a possible predominance of a matter
 over an antimatter.
 Naturally, what exactly the predominance of $u$- and $d$-quarks
 (from which the observed baryon matter is compounded, in ambient us
 areas of the Universe) probably indicates first of all on the
 predominance of the conforming flavors of right-handed polarized
 neutrinos with the enough high density of a degenerate Fermi --- gas.

 Let's mark, that we are not inclined to apply the adopted
 now classification of leptons and quarks on breeds,
 as massive unstable charged leptons it is possible
 correspond, contrary to the indicated classification, to
 $u$- and $d$-quarks, providing thereby (by large rest-masses)
 the stability of last ones (on the given interpretation us
 were pushed by the discovery of the fractional quantum Hall
 effect). Besides particles in the basic (vacuum)
 state (in our opinion) should have properties irrelevant to
 particles in excited state, owing to that the search of single
 quarks as free particles with a fractional charge is unpromising.
 Moreover the so-called quark confinement is connected to it ---
 the single quark becomes a particle in a basic state and it is
 admixed with weakly interacting particles in a vacuum, and a
 weakly interacting particle, which passes in an excited state
 with the corresponding properties (a chromatic charge, a fractional
 electric charge). from a vacuum takes its place in Hadron ---
 a strong interacting particles (baryon, meson).

     One of fundamental problems in physics is problem about
 the nature of rest-masses of fundamental particles, on which Mach
 has given the following answer: the inert properties of a body
 are determined by it interaction with all other bodies of the
 Universe, let even they are enough remote. It has allowed to
 Hoyle and Narlikar~\cite{hn} to consider a capability of an
 explanation of red displacements of electromagnetic radiations of
 remote galaxies and quasars at the expense of a change masses of
 fundamental particles. Now preferential the point of view, in
 which masses of fundamental particles are considered so, that they
 are induced by their interaction with hypothetical Higgs scalar
 fields, described by the disturbed symmetry~\cite{tay}. As the Higgs scalar
 particles till now are not found, it is possible to offer a
 hypothesis, in which masses of fundamental particles,
 belonging to the rapid subsystem, are induced by their interaction
 with particles of the slow subsystem (a dark matter~\cite{lei},
 a quintessence, etc.).
 In this connection we shall mark the huge value of the masses of
 the vector bosons $W^+, W^-, Z^o$, accountable for weak interaction,
 which one as against the massless photon, can interact with a
 background neutrinos directly. The given statement is easier to
 present through a Lagrangian (4), considering $M_4 $ by space
 Minkowski, and fields $\Phi_i^{(k)} (x), \Phi_{(j)}^l (x)$ by
 constants, owing to large density of weakly interacting particles
 and their homogeneous distribution in a space.
 Quadratic on fields~$A_{(i)}^{\underline{a}} (x)$ the summands in the
 Lagrangian (4) will be responsible for masses of particles, in
 addition the greatest masses will be to have those, which are
 quanta of fields entering in the expression (5) (owing to a large
 value of a constant $\kappa_1$ in a Lagrangian (4)).
 So if $\underline{r} =1$ and
\begin{equation}
\label{8}
 T_{\underline{a}}{}_{(k)}^{(i)}~\eta^{(j)(k)}+
 T_{\underline{a}}{}_{(k)}^{(j)}~\eta^{(i)(k)}
 = t_{\underline{a}}~\eta^{(i)(j)} ,
\end{equation}
 then the mass square of the vector boson being the quantum
 of field $A_i^{\underline{a}}$ has the form
\begin{equation}
\label{9}
 m^2 = 3 \kappa_1 t^2_{\underline{a}}/
 (\kappa_o \eta_{\underline{a}\underline{a}})
 - g^{jk} C_j{}_{\underline{a}}^{\underline{a}}
 C_k{}_{\underline{a}}^{\underline{a}} .
\end{equation}
 Let's remark, that in a case
 $C_i{}_{\underline{a}}^{\underline{a}}=0$
 the mass $m$ of a vector boson
 (which one under our supposition is the $Z$ boson)
 is determined under the formula (9)
 only by the value of the gravitational constant.

 So, masses of fundamental particles being excited states, will
 be determined by their interaction with particles from a ground
 state, and specially those, which  are present in the form of
 a Bose --- condensate from the coupled fermions, owing to,
 according to our opinion, its large density. The transition to
 a hot condition of the Universe was probably connected with a
 destruction of a Bose --- condensate and with an increase,
 accordingly, of a pressure of a Fermi --- gas.
 The given requirement causes us instead of a Lagrangian (7)
 to enter into a consideration the other Lagrangian,
 recording it as
\begin{equation}
\label{10}
 {\cal L}_t = {\cal L}(\Psi) +
 \kappa'_o {\cal F}_{ab}^c {\cal F}_{de}^f
 [\eta^{ad} (\delta_c^e \delta_f^b - 2 \delta_c^b \delta_f^e) +
 \eta^{be} (\delta^a_f \delta_c^d - 2 \delta^a_c \delta^d_f) +
 \eta_{cf} (\eta^{ad} \eta^{be} - 2 \eta^{ab} \eta^{de})]/4
\end{equation}
 ($a, b, c, d, e, f, g, h = 1, 2..., r \ge
 {\cal N}+{\underline{r}}$;
 $\kappa'_o$ is a constant one;
 $\eta_{ab}$ are metric tensor components of the flat space,
 and $\eta^{ab}$ are tensor components of a converse to basic one).
 In addition  intensities ${\cal F}_{ab}^c(B)$ of
 the bosons (gauge) fields ${\cal B}_a^c(B)$ will look like
\begin{equation}
\label{11}
 {\cal F}_{ab}^c = {\cal S}_d^c~(\Pi_a^i~\partial_i {\cal B}_b^d -
 \Pi_b^i~\partial_i {\cal B}_a^d + {\cal S}_{ab}^d),
\end{equation}
 where
\begin{equation}
\label{12}
     \Pi^i_a = {\cal B}_a^b~\xi_b^i, \quad
 {\cal S}_b^c = \delta_b^c - \xi_b^i~\Pi_i^d~({\cal B}_d^c -
 \beta_d^c), \quad
 {\cal S}_{ad}^b = ({\cal B}_a^c~T_c{}^e_d
 - {\cal B}_d^c~T_c{}^e_a)~{\cal B}_e^b -
 {\cal B}_a^c~{\cal B}_d^e~C_{ce}^b
\end{equation}
 (fields $\xi_a^i(x)$ determine a differential of a projection
 $d\pi$ from $\Omega_r \subset M_r$ in $\Omega_4
 \subset M_4$).
 We shall consider, that among fields ${\cal B}$ there are
 the mixtures $\Pi_a^i$ with non-zero vacuum averages $h_a^i$,
 and a selection of fields $\Pi_i^a(å)$, $\beta_c^a$
 are limited to the relations:
\begin{equation}
\label{13}
 \Pi_j^a~\Pi_a^i = \delta_j^i , \quad
 \beta_c^a~\xi_a^i = h_c^i
\end{equation}
 ($\delta_j^i$, $\delta_a^b$ are Kronecker deltas).

 The given Lagrangian is a most suitable one at the description
 of the hot stage of the Universe evolution
 because it is most symmetrical one
 concerning intensities of the gauge fields ${\cal F}_{ab}^c$.
 What is more, we shall demand, that the transition operators
 $T_a{}_á^b$ generate the symmetry, which follows from correlations:
\begin{equation}
\label{14}
 T_a{}_á^b~\eta^{cd} + T_a{}_c^d~\eta^{cb} = 0.
\end{equation}

 In absence of fields $\Pi_a^i(x)$ and $\Psi(x)$ at earlier stage
 of the Universe evolution
 the Lagrangian (10) becomes even more
 symmetrical (${\cal L}_t\propto {\cal B}^4$),
 so that the formation of fermions
 (the appearance of fields $\Psi$
 in a full Lagrangian ${\cal L}_t$)
 from primary bosons is a necessary condition (though it's
 not a sufficient one) of the transition of the Universe
 to the modern stage of its development by a spontaneous
 symmetry breaking.
 Only a formation of a Bose condensate
 from pairs of some class of
 fermions (the neutrino of different flavors)
 has resulted in a noticeable growth of
 rest--masses of those vector bosons ($W^+, W^-, Z^o$),
 which one interact with this class of fermions.
 In parallel there could be a growth of rest--masses
 of other fundamental particles,
 though and not all
 (photon, not interacting directly with neutrinos,
 has not a rest-mass).

 In addition the
 rest-masses, induced by the interaction with Bose --- condensate,
 of bosons $W^+, W^-, Z^o$ have decreased so, that the weak
 interaction has ceased to be gentle and all (or almost all)
 particles from the basic (vacuum) state steel to participate
 in an installation of a thermodynamic equilibrium. It also became
 the cause of an apparent increase of a density of particles.
 The reconversion to equilibrium state,
 bound with a formation and an increase of Bose --- condensate and
 described by an increase of an entropy,
 has resulted in effects,
 which was interpreted as a dilating of Universe.
 Guessing,
 that a density~$\rho$ of particles did not vary by this,
 and the script of a hot model of
 the Universe evolution is correct in general,
 we come to an estimation $\rho\sim m^3_{\pi} \sim 10^{-3} GeV^{3}$
 ($m_{\pi}$ is a mass of $\pi$-meson).

 As the gravitational interaction energy $\epsilon$
 of two macroscopic bodies (spaced $r$ apart)
 is directly proportional one to a particles number,
 forming given bodies and interacting with background neutrinos
 (the density $\rho_{\nu}$ of which),and also to a quasi-particles
 number (of which are exchanged every two particles of this bodies),
 then we have
\begin{equation}
\label{15}
 - \epsilon \sim (m_1 \sigma_{\nu})~(m_2 \sigma_{\nu})~
 \rho_{\nu} \sum_{k=1}^{\infty}
 e^{-k\theta r\sigma_{\nu} \rho_{\nu}} =
 \frac{m_1~m_2~\sigma_{\nu}^2~\rho_{\nu}}
 {e^{\theta r\sigma_{\nu}\rho_{\nu}} - 1}
 \approx
 \frac{m_1~m_2~\sigma_{\nu}}{r~\theta}
\end{equation}
 ($\sigma_{\nu}$ is the scattering cross--section of a neutrino on
 a particle of a macroscopic body, $\theta$ is a constant).
 This outcome allows to give an explanation to a known ratio~\cite{wei}
 $ H/G_N\approx m^3_{\pi}$,
 if to consider,
 that the Hubble constant~$H$ gives an estimation~$1/H$ of
 a length~$l\sim 1/(\rho\sigma_{\nu})$ of a free run of a particle
 in a vacuum
 at a modern stage of the Universe evolution,
 and to take into account the estimation
 of the gravitational constant~$G_N$
 ($G_N\sim\sigma_{\nu}$, the formula (15)).

\end {document}